\documentstyle[12pt,epsfig]{article}
\def\lapprox{\lower .7ex\hbox{$\;\stackrel{\textstyle <}{\sim}\;$}}
\def\d{{\rm d}}

\begin{document}
\begin{titlepage}
\vspace*{-1cm}
\begin{flushright}
DTP/95/80   \\
October 1995 \\
\end{flushright}
\vskip 3.cm
\begin{center}
{\Large\bf Production of $J/\psi$-pairs at HERA-$\vec{{\rm N}}$}
\vskip 1.cm
{\large  T.~Gehrmann}
\vskip .4cm
{\it Department of Physics, University of Durham \\
Durham DH1 3LE, England }\\
\vskip 3cm
\end{center}
\begin{abstract}
The production of $J/\psi$-pairs as a possible measure of the polarized
gluon distribution $\Delta G(x)$
is studied for proton--nucleon collisions at $\sqrt{s} =40
\;\mbox{GeV}^2$ (HERA-$\vec{{\rm N}}$).
 Possibilities of reconstructing the helicity
state of at least one of the $J/\psi$'s are critically reviewed.
The observation of production asymmetries in the single polarized mode
of HERA-$\vec{{\rm N}}$ is found to be not feasible.
\end{abstract}
\vfill
\end{titlepage}
\newpage

\section{Introduction}

So far, all experimental measurements of the spin structure of the
nucleon are measurements of the polarized structure function
$g_1(x,Q^2)$
\cite{exp}. At
leading order, this structure function reflects the polarization of quarks
in the nucleon
\begin{equation}
g_1(x,Q^2)=\frac{1}{2} \sum_q e_q^2 \left( q_{\uparrow} (x,Q^2) -
q_{\downarrow}
(x,Q^2) \right),
\end{equation}
while contributions from the polarization of gluons in the nucleon only enter
via higher order corrections to the above expression.
The data obtained from these experiments determine the polarization of
valence quarks
in the nucleon with an uncertainty of about $\pm$25\%.
 The polarization
of gluons in the nucleon, parametrized by
\begin{equation}
\Delta G(x,Q^2) = G_{\uparrow}(x,Q^2) - G_{\downarrow}(x,Q^2)
\end{equation}
can hardly be deduced from the $g_1$ measurements, although some estimates
exist\cite{polglue,grvpol,gspol}.
In the recent past, various experimental groups have proposed direct
measurements
of $\Delta G(x,Q^2)$ from open charm production in polarized lepton-nucleon
scattering \cite{hmc}, photoproduction of $J/\psi$ mesons \cite{breton} and
asymmetry measurements in polarized hadron-hadron collisions
\cite{rhic}. At this time,
only one of these experiments is approved \cite{rhic} and scheduled
for  data taking
in the year 2000.

The HERA-$\vec{{\rm N}}$ experiment\cite{nowak} would provide an
earlier opportunity
of singly polarized (unpolarized beam on polarized target) hadron--hadron
collisions. This experiment would use the (modified) HERMES target and
spectrometer
\cite{hermes} in the HERA proton beam, and hence the maximal
centre-of-mass energy
would be
\begin{equation}
\sqrt{s} = 40 \;\mbox{GeV}.
\end{equation}
Amongst various possibilities for the integrated luminosity, only the
so--called
high--luminosity option
\begin{equation}
\int {\cal L} \d t = 240 \;\mbox{pb}^{-1}
\end{equation}
will provide sufficient statistics for rare channels such as the one
under  consideration
in this paper.

In this paper, we will examine the possibility of determining $\Delta
G(x,Q^2)$  at
HERA-$\vec{{\rm N}}$  from a measurement of production asymmetries of $J/\psi$
pairs. In the
following section, we briefly outline the predictions of the colour
singlet  model
for the production of $J/\psi$ pairs. Section 3 examines possible ways
of  reconstructing
$J/\psi$ mesons and their helicity state at
HERA-$\vec{{\rm N}}$. In section 4, we compare
the production asymmetries in this channel for various, equally
possible  parametrizations
of $\Delta G(x,Q^2)$. Finally, section 5 contains our conclusions.

\section{Production of $J/\psi$ pairs in the colour singlet model}
The nonrelativistic colour singlet model \cite{cs} describes the production
of $J/\psi$ mesons as
arising from the
production of a charm quark pair. Both quarks
forming the $J/\psi$ have exactly half its velocity and form a colour
singlet. The transition amplitude from the quark pair to the meson state
is then inferred from the magnitude of the $J/\psi$ wavefunction at the
origin. This model sucessfully describes the production of $J/\psi$ mesons
in lepton-hadron collisions \cite{emc}, provided the theory prediction
is scaled with a $K$-factor accounting for higher order corrections. This
$K$-factor is of order 4 and almost independent of the kinematical variables.
The next-to-leading order calculation of $J/\psi$ photoproduction~\cite{zer}
is in better agreement with the experimental data, but still needs to be
scaled by a (smaller) $K$-factor.

This model can be generalized to the production of $J/\psi$ pairs in
hadron-hadron collisions, yielding the parton level cross section
\begin{equation}
\frac{\d \hat{\sigma} (a+b \rightarrow J/\psi J/\psi )}{\d \cos \Theta^{\star}}
= \frac{\pi^3\alpha_s^4 (1-M_{J/\psi}^2/\hat{s})^{1/2}}{72 \hat{s}} \frac{
\mid\! \Psi(0)\!\mid^4}{M_{J/\psi}^2} \frac{1}{64} \mid\! {\cal M} \!\mid^2
\qquad (a+b = q +\bar{q} \mbox{ or } g +g).
\end{equation}
The helicity averaged matrix element ${\cal M}$ was first calculated in
 \cite{hum}, and its full helicity dependence was derived in
\cite{bara}\footnote{
We would like to thank Sergey Baranov for providing the FORTRAN-code
for ${\cal  M}$}.

To relate this parton level cross section to a measurable observable
at the
hadronic level, it has to be convoluted with the corresponding parton
distributions
\begin{equation}
\d \sigma(p + p \rightarrow J/\psi J/\psi) = \sum_{a,b} \int \d x_1 \d x_2
f_{a/p} (x_1) f_{b/p} (x_2) \d \hat{\sigma} (a+b \rightarrow J/\psi J/\psi) .
\end{equation}
The above expression is dominated by the ($g+g$) subprocess, whose
contribution is about five times as big as the one from $(q+\bar{q})$ at
HERA-$\vec{{\rm N}}$ energies. In all further studies, we will
restrict us to the gluonic contribution.

To estimate the expected cross section at $\sqrt{s}= 40\;\mbox{GeV}$, we
have evaluated the above expression for the leading order parton distributions
given in \cite{ow} (DO1.1) and \cite{grvlo} (GRV94). The resulting $p_T^2$
distributions are shown in Figure 1. We use $\Lambda_{QCD}^{(4)} = 200
\;\mbox{MeV}$  and
$\mid\! \Psi(0) \!\mid^2 = 0.043\;\mbox{GeV}^3$, obtained from the
leptonic width of the $J/\psi$~\cite{pdg}.

The large discrepancy between both predictions reflects mainly the
uncertainty  on
the unpolarized gluon distribution $G(x)$ in the high-$x$ region.
For consistency with the polarized distributions considered in the
remainder  of
this paper, we will work with $G(x)$ from GRV94~\cite{grvlo}. For the proposed
high-luminosity option
of HERA-$\vec{{\rm N}}$, one can expect 1200 $J/\psi$ pairs to be produced.

\begin{table}[htb]
\begin{center}
\begin{tabular}{|l|c|c|} \hline
\rule[-1.2ex]{0mm}{4ex}& $\sigma^{tot}$ &
$\sigma(p_T^2>1\;\mbox{GeV}^2)$\\  \hline
\rule[-1.2ex]{0mm}{4ex} DO1.1 & $10.3 \;\mbox{pb}$ & $ 6.5 \;
\mbox{pb} $\\  \hline
\rule[-1.2ex]{0mm}{4ex} GRV94 & $4.9 \;\mbox{pb}$ & $ 3.0 \; \mbox{pb}
$\\  \hline
\end{tabular}
\caption{Total cross sections for $p p \rightarrow J/\psi J/\psi$
evaluated  for the
parton distributions of \protect{\cite{ow}} and \protect{\cite{grvlo}}.}
\end{center}
\end{table}

\begin{figure}[htb]
\begin{center}
{}~ \epsfig{file=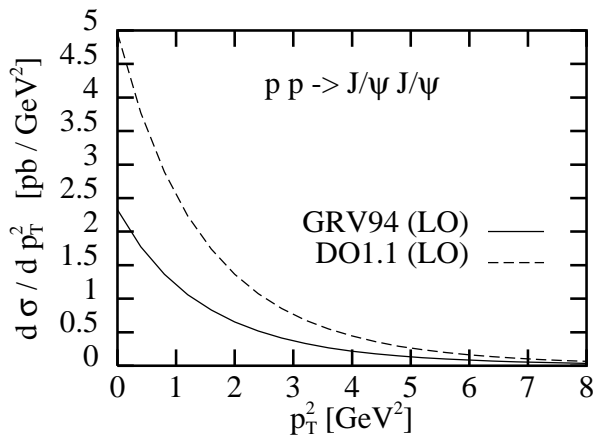,width=8cm}
\caption{Cross section for production of $J/\psi$ pairs at
HERA-$\vec{{\rm N}}$}
\end{center}
\end{figure}

It should be kept in mind that the colour singlet model assumes the colour
neutrality of the $(c\bar{c})$-pair to be obtained by a single, hard gluon
emision. This condition is only satisfied for sufficiently large
transverse momentum
of the final state particles. In the forward region $(p_T^2 \lapprox 0.5
\; \mbox{GeV}^2)$, the same neutral state can be obtained by the
multiple emission
of soft gluons. Hence the colour singlet model tends to underestimate the
cross section in this region.

\section{Reconstruction of the $J/\psi$}
The total cross section for proton-proton collisions at
$\sqrt{s} = 40\;\mbox{GeV}$ is
$\sigma^{tot}(pp)=41\;\mbox{mb}$ \cite{pdg}, ten
orders of magnitude bigger than the cross section for the production of
$J/\psi$ pairs. In order to identify these events in the background
of mulithadron production, a clear decay signature of at least one of the
$J/\psi$ mesons is needed. Only the leptonic decay $J/\psi \rightarrow
\gamma^{*} \rightarrow l^+ l^-$ can provide such a clear signature, as
the pair of oppositely charged leptons can be easily distinguished from
the hadronic background.
The branching ratio \cite{pdg}
\begin{equation}
Br ( J/\psi \rightarrow  \gamma^{*} \rightarrow l^+ l^-) = 12 \%
\end{equation}
of this decay channel therefore reduces the number of visible events.

HERA-$\vec{{\rm N}}$ will (at least for the first years of running)
only have a
polarized target, with an unpolarized beam. Information on the initial state
polarization will
therefore
have to be extracted from the final state. In the case
of $J/\psi$ pair production, at least the helicity of one of the $J/\psi$'s
has to be measured. As the $J/\psi$ is a massive spin-1 vector
meson, it has three possible helicity states: $-1,0,+1$. The $0$ and $\pm 1$
states correspond to different partial waves, and can therefore be easily
distinguished from the energy spectra of the decay products. Unfortunately,
no information on the initial state polarization can be gained from the $0$
state, as
the corresponding
 differential cross section is symmetric under the change of one
initial state helicity. We will discuss the possible decay channels of
the $J/\psi$ with a view to distinguishing the $+1$ and $-1$ helicity
states:
\begin{itemize}
\item[{(i)}] {\bf weak decays:} Parity violating weak decay modes could
provide a clear separation between these two states. As the $J/\psi$ does
not have any known weak decay modes, this possibility is ruled out.
\item[{(ii)}] {\bf leptonic decays:} Parity invariance of the electromagnetic
interaction relates the decay cross sections of both helicity states. As the
lepton helicities cannot be measured, both states are
indistinguishable.
\item[{(iii)}] {\bf decays to scalar mesons:} The distribution of the final
state particles in these decays is given by the $l=1, m=\pm 1$ partial waves.
As the partial waves for $m=-1$ and $m=1$ are identical for vector particles,
this decay channel cannot distinguish between these states.
\item[{(iv)}] {\bf radiative decays:} If the $J/\psi$ decays into a real photon
and scalar mesons (e.g. $J/\psi \rightarrow \eta_c \gamma$), the helicity
of the $J/\psi$ could be reconstructed from the measured helicity of
the photon\footnote{Even though such a measurement could be possible
in principle, it seems rather doubtful that it could be carried out with the
HERA-$\vec{{\rm N}}$ apparatus.}. This decay channel contributes with
a branching ratio
of about~4\%\cite{pdg}. Provided a helicity measurement on the photon,
this is the
only channel in which the helicity of the $J/\psi$ can be measured.
\end{itemize}

{}From the above considerations, it becomes clear that
a $J/\psi$ pair produced in single polarized proton-proton collisions can
only be used for an asymmetry measurement for the specific final state
configuration in which one $J/\psi$  decays leptonically while the other
decays into a photon accompanied by scalar mesons. The probability of
this configuration is
\begin{equation}
P = 2 \times \left(Br(J/\psi \rightarrow l^+l^-) \right) \times
\left(Br(J/\psi \rightarrow \gamma + \mbox{scalars}) \right) \simeq 1 \%.
\end{equation}
Therefore, only twelve of the expected 1200 events can provide an asymmetry
measurement under ideal experimental conditions at HERA-$\vec{{\rm N}}$.
It should therefore be already clear at this point that
such a measurement will fail to provide information on $\Delta G(x)$.
Regardless of this negative result, we will provide an estimate of the
asymmetries one could expect at HERA-$\vec{{\rm N}}$.

\section{Asymmetries at HERA-$\vec{{\rm N}}$ }
Under ideal experimental conditions at HERA-$\vec{{\rm N}}$, the spin
of the target proton
and the helicity of one of the two $J/\psi$ mesons can be measured in a
rather small fraction of the events. Using this information, we can construct
the following asymmetry
\begin{equation}
A= \frac{\d \sigma(p^{+} J/\psi^+)  + \d \sigma(p^{-} J/\psi^-)
  - \d \sigma(p^{-} J/\psi^+) - \d \sigma(p^{+} J/\psi^-) }{
 \d \sigma(p^{+} J/\psi^{\pm} ) + \d \sigma(p^{-} J/\psi^{\pm}) }.
\end{equation}
This asymmetry can be related to the parton level cross sections,
keeping in mind that the helicity state of the second $J/\psi$ is summed
over. For convenience, we use the following shorthand notation for
the parton level matrix elements of particular helicity combinations:
\begin{eqnarray}
\lefteqn{[h(g,\mbox{beam}) \; h(g,\mbox{target}) \; h(J/\psi_1) \;
h(J/\psi_2)] \equiv} \nonumber \\
& & \d \hat{\sigma} (
g(h(\mbox{beam})) + g(h(\mbox{target})) \rightarrow J/\psi(h_1)
J/\psi(h_2) ).
\end{eqnarray}
Omitting terms related by parity invariance, the asymmetry is
\begin{equation}
A = \frac{\int \d x_1 \d x_2 G(x_1,Q^2) \Delta G(x_2,Q^2) \left\{ 2
[\Sigma\Delta ++]  +
[\Sigma \Delta 0+] + [\Sigma\Delta +0] \right\} }{
\int \d x_1 \d x_2 G(x_1,Q^2) G(x_2,Q^2) [\Sigma \Sigma \pm \pm]},
\end{equation}
where $\Sigma$ ($\Delta$) denotes the sum (difference) of the
possible helicity states.

\begin{figure}[hb]
\begin{center}
{}~ \epsfig{file=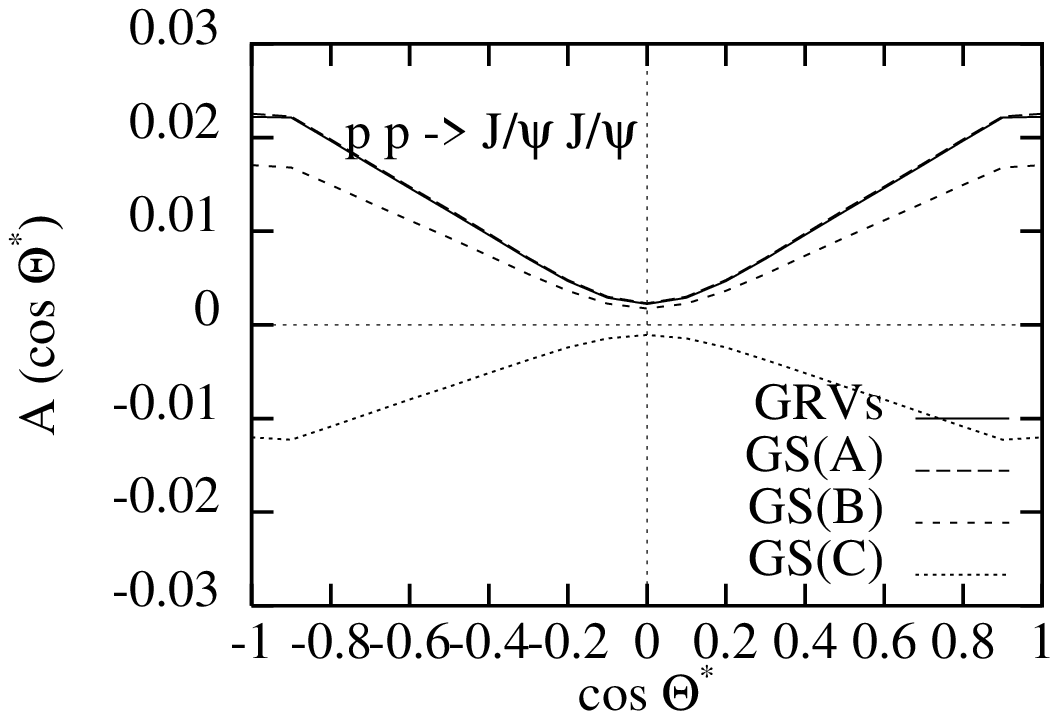,width=8cm}
\caption{Expected asymmerty in the single polarized mode of
HERA-$\vec{\rm N}$.}
\end{center}
\end{figure}

The scale of the parton distributions in the above expression and the
scale of $\alpha_s(Q^2)$ in the matrix elements is taken to be
$Q^2=(M_{J/\psi})^2$.
We have evaluated the above asymmetry as a function of the angle between the
$J/\psi$ pair and the proton beam direction in the parton-parton centre-of-mass
system (which can be reconstructed from the final state). The unpolarized
$G(x,Q^2)$ is taken from \cite{grvlo}. In Figure 2, we compare the
predictions obtained with the parametrizations of $\Delta G(x,Q^2)$ from
\cite{grvpol} (standard scenario) and \cite{gspol} (Gluon A-C).
Althogh the asymmetries obtained with these parametrizations are significantly
different from each other, $A(\cos \Theta^{\star})$ never exceeds 3\%.
The asymmetry becomes maximal if the $J/\psi$ pair is produced at very
small angles with respect to the proton beam, i.e. at low transverse momenta.

Keeping in mind the low number of reconstructable events, this small asymmetry
turns out to be unmeasurable in the single polarized mode of
HERA-$\vec{{\rm N}}$.
The situation would be different for a double polarized measurement (i.e. with
a polarized HERA proton beam): in this case, the reconstruction of helicities
in the final state is no longer necessary for an asymmetry
measurement.  Therefore,
one can expect about 270 $J/\psi$ pairs with at least one lepton pair
decay.  The
asymmetry can be defined in the standard way
\begin{equation}
A = \frac{\d \sigma (p^+ p^+) + \d \sigma (p^-p^-) - \d \sigma (p^+ p^-)
- \d \sigma (p^-p^+)}{ \d \sigma (pp)}.
\end{equation}
In terms of the parton densities this asymmetry reads
\begin{equation}
A = \frac{ \int \d x_1 \d x_2 \Delta G(x_1,Q^2) \Delta G(x_2,Q^2)
[+\Delta \Sigma\Sigma]
- [-\Delta \Sigma \Sigma]}{ \int \d x_1 \d x_2 G(x_1,Q^2) G(x_2,Q^2)
[\Sigma\Sigma\Sigma\Sigma]}
\end{equation}

\begin{figure}[hbpt]
\begin{center}
{}~ \epsfig{file=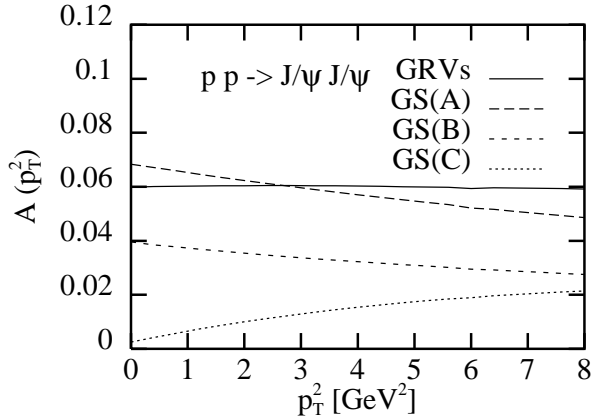,width=8cm}
\caption{Expected asymmetry in the double polarized mode of
HERA-$\vec{{\rm N}}$.}
\end{center}
\end{figure}
Figure 3 shows this asymmetry as a function of $p_T^2$ for the
parametrizations
of $\Delta G(x,Q^2)$ mentioned above. Depending on the
parametrization, the  asymmetry
could be as large as 7\% and only weakly depends on $p_T^2$. If
HERA-$\vec{{\rm N}}$ would
have a
polarized HERA proton beam available, this measurement could give some
indications
on $\Delta G(x,Q^2)$ for $x \approx 0.3$.

\section{Conclusions}
In this paper, we have studied the measurability of $\Delta G(x,Q^2)$ from the
production of $J/\psi$ pairs at HERA-$\vec{{\rm N}}$. We showed that a
measurement in
the single polarized mode is not feasible due to the low cross section and
the problematic reconstruction of the helicity state of the $J/\psi$.
At best,  one
could reconstruct twelve events in this channel, which is insufficient to
determine an asymmetry of at most 3\%. The situation improves in the
double  polarized
mode. In this mode, the production of $J/\psi$ pairs could provide
(amongst other
channels) a competitive measurement of $\Delta G(x,Q^2)$.

\section*{Acknowledgements}

\noindent Financial support from
the Gottlieb Daimler- und Karl Benz-Stiftung and the
Studienstiftung des deutschen Volkes is gratefully acknowledged.
This work was supported in part by the EU Programme
``Human Capital and Mobility'', Network ``Physics at High Energy
Colliders'', contract CHRX-CT93-0357 (DG 12 COMA).
\goodbreak

\end{document}